\documentclass[conference]{IEEEtran}
\IEEEoverridecommandlockouts
\input{sty.sty}
\usepackage{comment}
\usepackage{multirow}
\usepackage{cite}
\usepackage{amsmath,amssymb,amsfonts}
\usepackage{algorithmic}
\usepackage{graphicx}
\usepackage{textcomp}
\usepackage{xcolor}
\def\BibTeX{{\rm B\kern-.05em{\sc i\kern-.025em b}\kern-.08em
    T\kern-.1667em\lower.7ex\hbox{E}\kern-.125emX}}
\begin{document}

\title{RaftGP: Random Fast Graph Partitioning
}

\author{
\IEEEauthorblockN{Yu Gao\IEEEauthorrefmark{1}\IEEEauthorrefmark{2} Meng Qin\IEEEauthorrefmark{1}\IEEEauthorrefmark{2}\IEEEauthorrefmark{4} Yibin Ding\IEEEauthorrefmark{2} 
 Li Zeng\IEEEauthorrefmark{3}
 Chaorui Zhang\IEEEauthorrefmark{2}
 Weixi Zhang\IEEEauthorrefmark{2} 
 Wei Han\IEEEauthorrefmark{2} 
 Rongqian Zhao\IEEEauthorrefmark{3}
 Bo Bai\IEEEauthorrefmark{2}}
\IEEEauthorblockA{\IEEEauthorrefmark{2}Theory Lab, Central Research Institute, 2012 Labs, Huawei Technologies Co., Ltd.}
\IEEEauthorblockA{\IEEEauthorrefmark{3}Algorithm \& Technology Development Department, Global Technical Service Department, Huawei Technologies Co., Ltd.}
\IEEEauthorblockA{\IEEEauthorrefmark{4}Department of Computer Science \& Engineering, Hong Kong University of Science \& Technology}
\IEEEauthorblockA{\IEEEauthorrefmark{1}The first two authors contributed equally. Corresponding Author: Meng Qin (mengqin\_az@foxmail.com).}
\IEEEauthorblockA{(\textbf{Innovation Award Winner} of IEEE HPEC 2023 Graph Challenge: https://graphchallenge.mit.edu/champions)}
}

\maketitle

\begin{abstract}
Graph partitioning (GP), a.k.a. community detection, is a classic problem that divides the node set of a graph into densely-connected blocks. Following prior work on the IEEE HPEC Graph Challenge benchmark and recent advances in graph machine learning, we propose a novel RAndom FasT Graph Partitioning (RaftGP) method based on an efficient graph embedding scheme. It uses the Gaussian random projection to extract community-preserving features from classic GP objectives. These features are fed into a graph neural network (GNN) to derive low-dimensional node embeddings. Surprisingly, our experiments demonstrate that a randomly initialized GNN even without training is enough for RaftGP to derive informative community-preserving embeddings and support high-quality GP. To enable the derived embeddings to tackle GP, we introduce a hierarchical model selection algorithm that simultaneously determines the number of blocks and the corresponding GP result. We evaluate RaftGP on the Graph Challenge benchmark and compare the performance with five baselines, where our method can achieve a better trade-off between quality and efficiency. In particular, compared to the baseline algorithm \cite{tiago14} of the IEEE HPEC Graph Challenge, our method is 6.68x -- 23.9x faster on graphs with 1E3 -- 5E4 nodes and at least 64.5x faster on larger (1E5 node) graphs on which the baseline takes more than 1E4 seconds. Our method achieves better accuracy on all test cases. We also develop a new graph generator to address some limitations of the original generator in the benchmark.
\end{abstract}


\section{Introduction}\label{sec:introduction}
For many real-world complex systems (e.g., social, communication, and biological networks), graphs serve as a generic abstraction of system entities and their relations in terms of nodes and edges. Graph partitioning (GP), a.k.a. disjoint community detection, is a classic problem that divides the node set of a graph into disjoint blocks (communities) with dense linkages distinct from other blocks. The partitioning of blocks (communities) can reveal the structures and functions of a graph \cite{qin2018adaptive,li2019identifying,qin2021dual} while supporting advanced applications (e.g., parallel task assignment \cite{hendrickson2000graph}, wireless network decomposition \cite{dai2017optimal}, and traffic classification \cite{qin2019towards}).

GP on large graphs is usually challenging because it can be formulated as several NP-hard combinatorial optimization objectives.
Previous studies on the IEEE HPEC Graph Challenge benchmark \cite{DBLP:conf/hpec/KaoGHJKMMRSSSS17} have provided a series of solutions to alleviating the NP-hard challenge of GP.
Knyazev et al. \cite{zhuzhunashvili2017preconditioned} solved the eigen decomposition of graph Laplacians via locally optimal block preconditioned conjugate gradient and accelerated the spectral clustering algorithm.
Durbeck et al. \cite{durbeck2021dpgs} explored the ability of graph summarization techniques to preserve the underlying community structures of graphs.
Uppal et al. \cite{uppal2021faster} and Wanye et al. \cite{wanye2019fast} proposed a top-down divide-and-conquer algorithm and a sampling strategy to speed up the Bayes inference of the stochastic block model (SBM).

Different from the aforementioned studies that focus on the engineering improvements of existing GP techniques (e.g., spectral clustering and Bayesian SBM), we propose a novel RAndom FasT Graph Partitioning (RaftGP) method following recent advances in graph machine learning \cite{zhang2020deep}.
It adopts an efficient community-preserving graph embedding scheme including (\romannumeral1) random feature extraction, (\romannumeral2) random embedding derivation, and (\romannumeral3) hierarchical model selection.

We first use the efficient Gaussian random projection \cite{arriaga2006algorithmic} to extract two sets of low-dimensional features from classic GP objectives of normalized cut (NCut) minimization \cite{von2007tutorial} and modularity maximization \cite{newman2006modularity}, which induces two variants of RaftGP.
The extracted features are believed to encode the community structures of graphs because the random projection can preserve the geometric structures (e.g., in terms of distances) between inputs (i.e., components about graph structures in the two GP objectives) with rigorous theoretical guarantees \cite{arriaga2006algorithmic}.

To derive low-dimensional embeddings, we then feed the extracted features into a graph neural network (GNN), which further enhances their abilities to preserve community structures via multi-hop feature aggregation while reducing the dimensionality.
In contrast to existing GNN-based methods \cite{wilder2019end,chen2019supervised,qin2022trading} that require time-consuming training, we get inspiration from the efficient random projection and consider an extreme design of RaftGP, where we do not apply any training procedures to GNN.
Surprisingly, our experiments demonstrate that one feedforward propagation of a randomly initialized GNN , even without training, is enough for RaftGP to derive informative community-preserving embeddings.

To enable the embedding scheme to tackle GP, we introduce a hierarchical model selection algorithm that simultaneously determines the proper number of blocks and corresponding block membership (i.e., the GP result). It recursively partitions the node set into two subsets by applying $K$Means to the derived node embeddings, where a novel \textit{local modularity} metric is used to stop the recursion.

We follow the experiment settings of the IEEE HPEC Graph Challenge to evaluate RaftGP and compare the performance with five baselines. However, we notice some limitations of the original data generators in the benchmark.
To address these limitations, we develop a new algorithm that can generate graphs from the exact SBM distribution in nearly linear time.

\section{Problem Statements \& Preliminaries}\label{sec:background}
In this study, we consider undirected unweighted graphs. In general, a graph can be represented as $G = (V, E)$, where ${V} = \{ v_1, v_2, \cdots, v_N\}$ and ${E} = \{ (v_i, v_j) | v_i, v_j \in V\}$ are the sets of nodes and edges. The topology structure of a graph can be described by an adjacency matrix ${\AA} \in \{ 0, 1\}^{N \times N}$, where ${\AA}_{ij} = {\AA}_{ji} = 1$ if $(v_i, v_j) \in {E}$ and ${\AA}_{ij} = {\AA}_{ji} = 0$ otherwise.
We study the following \textit{graph partitioning} (GP) problem.
\begin{definition}[{\textbf{Graph Partitioning}}, GP]\label{def:GP}
Given a graph $\nocal{G}$, GP (a.k.a. disjoint community detection) aims to partition the node set $\nocal{V}$ into $K$ disjoint subsets (i.e., blocks or communities) $\nocal{C} = \{ \nocal{C}_1, \cdots , \nocal{C}_K\}$ such that (\romannumeral1) within each block the linkage is dense but (\romannumeral2) between different blocks the linkage is relatively loose. In this study, we assume that $K$ is not given and should be determined by a model selection procedure.
\end{definition}

Our RaftGP method follows a novel community-preserving embedding scheme based on some classic combinatorial optimization objectives of GP (e.g., NCut minimization \cite{von2007tutorial} and modularity maximization \cite{newman2006modularity}).

\begin{definition}[\textbf{Graph Embedding}]
Given a graph $\nocal{G}$, graph embedding (a.k.a. graph representation learning) learns a function $f: \nocal{V} \mapsto {\mathbb{R} ^d}$ that maps each node $v_i$ to a low-dimensional vector representation (a.k.a. embedding) ${\bf{z}}_i \in \mathbb{R}^{d}$ ($d \ll N$). The derived embeddings $\{ {\bf{z}}_i \}$ are expected to preserve some major properties (e.g., community structures) of graph topology. For instance, nodes $(v_i, v_j)$ with similar properties (e.g., in the same block) should have close embeddings $({\bf{z}}_i, {\bf{z}}_j)$ (e.g., in terms of distance). It is a unified framework that can support various graph inference tasks by combining $\{ {\bf{z}}_i \}$ with specific downstream modules. For GP, one can apply the $K$Means clustering algorithm to $\{ {\bf{z}}_i \}$, which derives the GP result $\nocal{C}$ w.r.t. the determined number of blocks $K$.
\end{definition}

\begin{definition}[\textbf{NCut Minimization}]
Given a graph $\nocal{G}$ and the number of blocks $K$ to be partitioned, NCut minimization aims to derive a GP result $\nocal{C}$ that minimizes the NCut metric:
\begin{equation}\label{Eq:NCut-Min}
    \mathop {\min }\limits_\nocal{C} ~{\mathop{\rm NCut}\nolimits} (\nocal{G}, K) \equiv \frac{1}{2}\sum\nolimits_{r = 1}^{{K}} {[{\mathop{\rm cut}\nolimits}(\nocal{C}_r,\nocal{\overline C}_r)/{\mathop{\rm vol}\nolimits} (\nocal{C}_r)]},
\end{equation}
where $\nocal{\overline C}_r = {{\nocal{V}}} - \nocal{C}_r$ denotes the complementary set of ${\nocal{C}_r}$; ${\mathop{\rm cut}\nolimits}(\nocal{C}_r,{\nocal{\overline C}_r}) = \sum\nolimits_{{v_i} \in \nocal{C}_r,{v_j} \in \nocal{\overline C}_r} {{ {\bf{A}}_{ij} }}$ is the cut between ${C_r}$ and ${{\overline C}_r}$; ${\mathop{\rm vol}\nolimits} (\nocal{C}_r) = \sum\nolimits_{{v_i} \in \nocal{C}_r, {v_j} \in \nocal{V}} {{ {\bf{A}}_{ij} }}$ is the volume of ${\nocal{C}_r}$. (\ref{Eq:NCut-Min}) can be further rewritten into the following matrix form:
\begin{equation}\label{Eq:NCut-Min-Mat}
\resizebox{.91\linewidth}{!}{$
\mathop {\min }\limits_{\bf{H}}~{\mathop{\rm tr}\nolimits} ({{\bf{H}}^T}{\bf{LH}}) {\rm{~s.t.~}} {{\bf{H}}_{ir}} = \left\{ {\begin{array}{*{20}{l}}
{ {[{{\deg}(v_i)}/{\mathop{\rm vol}\nolimits} ({\nocal{C}_r})]^{0.5}} ,{v_i} \in {\nocal{C}_r}}\\
{0,{\rm{~otherwise}}}
\end{array}} \right.,
$}
\end{equation}
where ${\bf{L}} \equiv {\bf{I}}_N - {\bf{D}}^{-0.5} {\bf{A}} {\bf{D}}^{-0.5}$ is the normalized Laplacian matrix of ${\bf{A}}$; ${\bf{D}} \equiv {\mathop{\rm diag}\nolimits} ({\deg} (v_1), \cdots ,\deg {v_N})$ is defined as a degree diagonal matrix with ${\deg} (v_i)$ as the degree of node $v_i$; ${\bf{I}}_N$ is an $N$-dimensional identity matrix; ${\bf{H}}$ consists of the scaled block membership indicators.
\end{definition}

\begin{definition}[\textbf{Modularity maximization}]
\label{def:modularity} 
For the given $\nocal{G}$ and $K$, modularity maximization aims to derive a GP result $\nocal{C}$ that maximizes the following modularity metric:
\begin{equation}\label{Eq:Mod-Max}
\resizebox{.91\linewidth}{!}{$
    \mathop {\max}\limits_\nocal{C} {\rm{Mod}}(\nocal{C}, K) \equiv \frac{1}{{2e}}\sum\limits_{r = 1}^K {\sum\limits_{{v_i},{v_j} \in {\nocal{C}_r}} {[{{\bf{A}}_{ij}} - \frac{{{{\deg} (v_i)}{{\deg} (v_j)}}}{{2e}}]} },
$}
\end{equation}
where ${\deg} (v_i)$ is the degree of node $v_i$; $e$ is the number of edges. Similarly, (\ref{Eq:Mod-Max}) can also be rewritten into a matrix form:
\begin{equation}\label{Eq:Mod-Max-Mat}
\mathop {\min }\limits_{\bf{H}} ~ - {\mathop{\rm tr}\nolimits} ({{\bf{H}}^T}{\bf{QH}}) {\rm{~s.t.~}} {{\bf{H}}_{ir}} = \left\{ {\begin{array}{*{20}{l}}
{1,~{v_i} \in {\nocal{C}_r}}\\
{0,{\rm{~otherwise}}}
\end{array}} \right.,
\end{equation}
where ${\bf{Q}} \in \mathbb{R}^{N \times N}$ is defined as the modularity matrix with ${\bf{Q}}_{ij} = {\bf{A}}_{ij} - {\deg} (v_i) {\deg} (v_j) / (2e)$; ${\bf{H}}$ consists of the block membership indicators.
\end{definition}

Although NCut minimization and modularity maximization are NP-hard and need a pre-set number of blocks $K$, our method does not directly solve these combinatorial optimization problems. Instead, we only focus on how to efficiently extract informative community-preserving features from (\ref{Eq:NCut-Min-Mat}) and (\ref{Eq:Mod-Max-Mat}), which is detailed later in Section~\ref{sec:feat_ext}.

The benchmarks of the IEEE HPEC Graph Challenge contain graphs drawn from the SBM with different settings. One baseline algorithm \cite{tiago14} is also based on maximizing the posterior probability of the SBM.
\begin{definition}[\textbf{Stochastic Block Model}, SBM]
\label{def:SBM}
Let $\boldsymbol{\theta} \in \mathbb{R}^{N}_+$ and ${\bf{c}} \in[K]^N$ be the degree correction vector and the block assignment vector. Let ${\bf{\Omega}} \in \mathbb{R}_+^{K \times K}$ be the block interaction matrix. The SBM independently generates an edge for each pair of nodes $(v_i, v_j)$ (i.e., each entry of adjacency matrix ${\bf{A}}$) via a Poisson distribution $\mathbf{A}_{ij}\sim Pois(\lambda_{ij} \equiv \boldsymbol{\theta}_i \boldsymbol{\theta}_j {\bf{\Omega}}_{c_i c_j})$.
\end{definition}

\section{Methodology}\label{sec:methodology}

To alleviate the NP-hard challenge of GP, we propose a novel RaftGP method following an efficient community-preserving embedding scheme. In this section, we elaborate on the three major procedures of RaftGP, which are (\romannumeral1) \textit{random feature extraction}, (\romannumeral2) \textit{random embedding derivation}, and (\romannumeral3) \textit{hierarchical model selection}.

\subsection{Random Feature Extraction}\label{sec:feat_ext}
We first extract informative features from the key components regarding graph structures in classic GP objectives.
For NCut minimization defined in (\ref{Eq:NCut-Min-Mat}), ${\bf{M}} \equiv {\bf{D}}^{-0.5} {\bf{A}} {\bf{D}}^{-0.5}$ in the Laplacian matrix ${\bf{L}}$ is used to describe the graph topology. For modularity maximization, the modularity matrix ${\bf{Q}}$ in (\ref{Eq:Mod-Max-Mat}) is the primary component regarding graph structures.

In particular, $\{ {\bf{M}}, {\bf{Q}} \}$ can be considered as a reweighting of the original neighbor structures described by the adjacency matrix ${\bf{A}}$, where the similarity between $({\bf{M}}_{i,:}, {\bf{M}}_{j,:})$ (or $({\bf{Q}}_{i,:}, {\bf{Q}}_{j,:})$) indicates the neighbor similarity between nodes $(v_i, v_j)$. In general, nodes $(v_i, v_j)$ with a higher neighbor similarity (i.e., denser local linkage) are more likely to be partitioned into the same block. Hence, we believe that $\{ {\bf{M}}, {\bf{Q}} \}$ encode key characteristics about the community structures of a graph.
Tian et al. \cite{tian2014learning} and Yang et al. \cite{yang2016modularity} have also validated the potential of ${\bf{M}}$ and ${\bf{Q}}$ to derive community-preserving embeddings via deep auto-encoders.
Instead of using auto-encoders, we propose a more efficient strategy to extract community-preserving features (denoted by ${\bf{Y}} \in \mathbb{R}^{N \times L}$) from $\{ {\bf{M}}, {\bf{Q}} \}$, with the feature dimensionality $L \ll N$.

Since most real-world graphs have sparse topology, ${\bf{M}}$ can be treated as a sparse matrix. However, ${\bf{Q}}$ is still a dense matrix. To fully utilize the sparsity of topology, we introduce a reduced modularity matrix $\bf{\tilde Q} \in \mathbb{R}^{N \times N}$, where ${\bf{\tilde Q}}_{ij} = {\bf{Q}}_{ij}$ if $(v_i, v_j) \in E$ and ${\bf{\tilde Q}}_{ij} = 0$ otherwise.
Although the reduction of ${\bf{Q}}$ may lose some information, our experiments demonstrated that ${\bf{\tilde Q}}$ is enough to derive informative community-preserving embeddings to support high-quality GP.

Let ${\bf{X}} \in \{ {\bf{M}}, {\bf{\tilde Q}} \}$ be the structural component from NCut minimization or modularity maximization, which induces two variants of RaftGP.
We extract the features ${\bf{Y}}$ via
\begin{equation}\label{Eq:Feat-Ext}
    {\bf{Y}} = {\bf{X}} {\bf{\Theta}} {\rm{~with~}} {\bf{\Theta}} \in \mathbb{R}^{N \times L}, {\bf{\Theta}}_{ir} \sim {\mathcal{N}}(0, L^{ - 0.5}),
\end{equation}
where ${\bf{Y}}_{i,:}$ is the extracted feature of node $v_i$. Concretely, (\ref{Eq:Feat-Ext}) is the Gaussian random projection \cite{arriaga2006algorithmic} of ${\bf{X}} \in \mathbb{R}^{N \times N}$, an efficient dimension reduction technique that can preserve the geometry structures (e.g., in terms of distances) of input features in finite-dimensional $l_1$ \cite{Bourgain1985embedding} and $l_2$ \cite{1984Extensions} spaces with rigorous theoretical guarantees. For instance, nodes $(v_i, v_j)$ with close $({\bf{X}}_{i,:}, {\bf{X}}_{j,:})$ are guaranteed to have close $({\bf{Y}}_{i,:}, {\bf{Y}}_{j,:})$.

Since ${\bf{X}}$ is sparse, the time complexity of random projection is $O(|E|L)$ by using the sparse-dense matrix multiplication. Given a graph ${G}$, the complexities to compute ${\bf{M}}$ and ${\bf{\tilde Q}}$ are both $O(|E|)$. In summary, the overall complexity of the random feature extraction step is $O(|E|L)$.

\subsection{Random Embedding Derivation}\label{sec:emb-der}
Although the extracted features (described by ${\bf{Y}} \in \mathbb{R}^{N \times L}$) have the initial ability to encode characteristics about community structures, we derive the final community-preserving embeddings (denoted by ${\bf{Z}} \in \mathbb{R}^{N \times d}$) by feeding ${\bf{Y}}$ into a multi-layer GNN, which further enhances the extracted features via the feature aggregation of GNN while reducing the feature dimensionality to $d < L$. We adopt GCN \cite{kipf2016semi} to build the multi-layer GNN because it is easy to implement and has a low complexity of feature aggregation.

Let ${\bf{Z}}^{(k-1)} \in \mathbb{R}^{N \times L_{k-1}}$ and ${\bf{Z}}^{(k)} \in \mathbb{R}^{N \times L_{k}}$ be the input and output of the $k$-th GNN layer, where ${\bf{Z}}^{(0)} = {\bf{Y}}$; $L_{k-1}$ and $L_k$ are the input and output feature dimensionality. The $k$-th GNN layer can be described as 
\begin{equation}\label{Eq:GCN}
    {{\bf{Z}}^{(k)}} = {f_{{\rm{act}}}}({{{\bf{\hat D}}}^{ - 0.5}}{\bf{\hat A}}{{{\bf{\hat D}}}^{ - 0.5}}{{\bf{Z}}^{(k - 1)}}{{\bf{W}}^{(k)}}),
\end{equation}
where ${\bf{Z}}_{i,:}^{(k)}$ is the (intermediate) representation of node $v_i$; $f_{\rm{act}} (\cdot)$ is an activation function to be specified (e.g., tanh in our implementation); ${\bf{\hat A}} \equiv {\bf{A}} + {\bf{I}}_N$ is the adjacency matrix with self-edges; ${\bf{\hat D}}$ is the degree diagonal matrix of ${\bf{\hat A}}$; ${\bf{W}}^{(k)} \in \mathbb{R}^{L_{k-1} \times L_{k}}$ is a trainable parameter. In particular, ${\bf{Z}}_{i,:}^{(k)}$ is the non-linear aggregation (i.e., weighted mean) of features w.r.t. $\{ v_i\} \cup {\mathop{\rm N}\nolimits} ({v_i})$ from the previous layer, with ${\mathop{\rm N}\nolimits} ({v_i})$ as the neighbor set of $v_i$.
The feature aggregation forces nodes $(v_i, v_j)$ with similar neighbor sets $({\mathop{\rm N}\nolimits} ({v_i}), {\mathop{\rm N}\nolimits} ({v_j}))$ (i.e., dense local linkage)  to have similar features $({\bf{Z}}_{i,:}^{(k)}, {\bf{Z}}_{j,:}^{(k)})$, thus enhancing the ability to encode characteristics regarding community structures.
The multi-layer structure can further extend the feature aggregation to the local topology induced by multi-hop neighbors.

Before feeding ${\bf{Z}}^{(k)}$ into the next GNN layer, we recommend conducting the row-wise $l_2$-normalization via ${\bf{Z}}_{i,:}^{(k)} \leftarrow {\bf{Z}}_{i,:}^{(k)}/|{\bf{Z}}_{i,:}^{(k)}{|_2}$, which controls the scale of the (intermediate) representation of each node $v_i$.
We treat the normalized representations from the last GNN layer (denoted by ${\bf{Z}} \in \mathbb{R}^{N \times d}$) as the final community-preserving embeddings. In the rest of this paper, we use ${\bf{z}}_i \in \mathbb{R}^{d}$ to represent the embedding of $v_i$.

The powerful inference abilities of some state-of-the-art GNN-based GP methods (e.g., \textit{ClusterNet} \cite{wilder2019end}, \textit{LGNN} \cite{chen2019supervised}, and \textit{ICD} \cite{qin2022trading}) rely on the offline training of model parameters (e.g., $\{ {\bf{W}}^{(k)}\}$) via the time-consuming gradient descent.

Inspired by the efficient random projection described in (\ref{Eq:Feat-Ext}), we consider an extreme design of RaftGP, where we do not apply any training procedures to the multi-layer GNN. Our experiments demonstrate that one feedforward propagation through a randomly initialized GNN without any training is enough for RaftGP to derive informative community-preserving embeddings and support high-quality GP.
A reasonable interpretation is that each GNN layer described in (\ref{Eq:GCN}) can be considered as a special random projection similar to (\ref{Eq:Feat-Ext}). In the $k$-th layer, the geometric properties (e.g., relative distances) between the aggregated representations (i.e., ${{{\bf{\hat D}}}^{ - 0.5}}{\bf{\hat A}}{{{\bf{\hat D}}}^{ - 0.5}}{{\bf{Z}}^{(k - 1)}}$) can be preserved with theoretical guarantees, even though it is multiplied by a random matrix ${\bf{W}}^{(k)}$ and mapped to another space with lower dimensionality. The nonlinear activation and $l_2$-normalization will also not largely change the geometric properties.

For the GNN feature aggregation in (\ref{Eq:GCN}), one can treat ${\bf{\hat D}}^{-0.5} {\bf{\hat A}} {\bf{\hat D}}^{-0.5}$ as a sparse matrix and use the efficient sparse-dense matrix multiplication for implementation. In this setting, the complexity of GNN feedforward propagation is $O(|E|L + NL{L_1} + |E|{L_1} + N{L_1}{L_2} + \cdots) = O(|E|L + NL{L_1})$, where we usually set $L_k > L_{k+1}$. Assume that $\{ L, L_1\}$ have the same magnitude with the embedding dimensionality $d$. The complexity of the embedding derivation step is $O(|E|d + Nd^2)$.

\subsection{Hierarchical Model Selection}\label{sec:recursion}

Graph embedding is a unified framework that can support various inference tasks by combining the derived embeddings $\{ {\bf{z}}_i \}$ with specific downstream modules. To enable RaftGP to support the GP task stated in Definition~\ref{def:GP}, the model selection problem (i.e., determining a proper number of blocks $K$) is critical.
We introduce an efficient hierarchical model selection algorithm based on a novel \textit{local modularity} metric.

\begin{definition}[\textbf{Local Modularity}]\label{def:local_modularity}
Let $U \subseteq V$ be a subset of nodes in graph $G = (V, E)$. Let $\deg_G(v)$ be the degree of node $v$ in $G$.
Suppose $U$ is partitioned into $K$ disjoint blocks $({U_1}, \cdots ,{U_K})$. We define the local modularity metric as
\begin{equation}\label{eq:lcl-mod}
    {\mathop{\rm LMod}\nolimits} ({U_1}, \cdots ,{U_K}; G) \equiv \sum\nolimits_{r = 1}^K {[\frac{{{m_r}}}{e} - {{(\frac{{{\overline m_r}}}{{2 e}})^2 }}]},
\end{equation}
where ${m_r}$ is the number of intra-block edges of $U_r$; ${{\overline m}_r} = \sum\nolimits_{v \in {U_r}} {{{\deg }_G}(v)}$ is total degree of nodes in $U_r$; $2e = \sum\nolimits_{r = 1}^K {{\overline m_r}}$ is the total degree of all nodes in $U$.
\end{definition}

Different from the standard modularity on $G[U]$ which considers only edges induced by $U$, local modularity also takes account of the edges from $U$ to $V - U$. When partitioning a subgraph, local modularity keeps its interaction with the remaining graph and thus ensures that we do not further partition a true block into sub-blocks.

\begin{algorithm2e}[t]\small
\caption{\small Hierarchical Model Selection}
\label[Algorithm]{alg:recursion}
\SetKwProg{Proc}{procedure}{}{}
\SetKwProg{Globals}{global variables}{}{}
\Globals{}{
Graph $G = (V, E)$; 
Derived node embeddings $\{ {\bf{z}}_i \}$; Current GP result $C$ (with $C \leftarrow \emptyset$ at the beginning)\\}
\Proc{$\textsc{Partition}(U, G)$}{
    $m_1\leftarrow {\mathop{\rm LMod}\nolimits}(U; G)$ via (\ref{eq:lcl-mod}) \label{line:singleblock}\\
    Apply $K$Means to $\{ {\bf{z}}_i | v_i \in U \}$ with $K=2$ and get two subsets $\{ U_1, U_2 \}$\label{line:partitionintotwo}\\
    $m_2\leftarrow {\mathop{\rm LMod}\nolimits}(U_1, U_2; G)$ via (\ref{eq:lcl-mod})\label{line:twoblocks}\\
    \If{$|U_1| \le \epsilon$ {\bf{or}} $|U_1| \le \epsilon$ {\bf{or}} $m_1>m_2$\label{line:threshold}} {
        Add $U$ as a new block to $C$\label{line:newblock}
    } \Else {
        $\textsc{Partition}(U_1, G)$\label{line:recursion1}\\
        $\textsc{Partition}(U_2, G)$\label{line:recursion2}\\
    }
}
\end{algorithm2e}

\begin{figure}
\begin{tikzpicture}[scale=0.24]
	\begin{pgfonlayer}{nodelayer}
		\node [style=none] (14) at (-6.5, 3.75) {};
		\node [style=none] (15) at (2, 3.75) {};
		\node [style=none] (16) at (2, -4) {};
		\node [style=none] (17) at (-6.5, -4) {};
		\node [style=new style 0] (18) at (-5.25, 1.75) {};
		\node [style=new style 0] (19) at (-3.5, 2.5) {};
		\node [style=new style 0] (20) at (-4, 1.25) {};
		\node [style=new style 0] (21) at (-5.25, -2.75) {};
		\node [style=new style 0] (22) at (-3.75, -1.75) {};
		\node [style=new style 0] (23) at (-3.5, -3) {};
		\node [style=new style 0] (24) at (-1, 1.5) {};
		\node [style=new style 0] (25) at (0, 2.5) {};
		\node [style=new style 0] (26) at (1, 2.5) {};
		\node [style=new style 0] (27) at (-1.25, -0.5) {};
		\node [style=new style 0] (28) at (-0.25, -1) {};
		\node [style=new style 0] (29) at (0.75, -1.25) {};
		\node [style=new style 0] (30) at (-1.5, -2.5) {};
		\node [style=new style 0] (31) at (-1.25, -3.25) {};
		\node [style=none] (32) at (3, 3.75) {};
		\node [style=none] (33) at (11.5, 3.75) {};
		\node [style=none] (34) at (11.5, -4) {};
		\node [style=none] (35) at (3, -4) {};
		\node [style=new style 0] (36) at (4.25, 1.75) {};
		\node [style=new style 0] (37) at (6, 2.5) {};
		\node [style=new style 0] (38) at (5.5, 1.25) {};
		\node [style=new style 0] (39) at (4.25, -2.75) {};
		\node [style=new style 0] (40) at (5.75, -1.75) {};
		\node [style=new style 0] (41) at (6, -3) {};
		\node [style=new style 0] (42) at (8.5, 1.5) {};
		\node [style=new style 0] (43) at (9.5, 2.5) {};
		\node [style=new style 0] (44) at (10.5, 2.5) {};
		\node [style=new style 0] (45) at (8.25, -0.5) {};
		\node [style=new style 0] (46) at (9.25, -1) {};
		\node [style=new style 0] (47) at (10.25, -1.25) {};
		\node [style=new style 0] (48) at (8, -2.5) {};
		\node [style=new style 0] (49) at (8.25, -3.25) {};
		\node [style=none] (50) at (12.5, 3.75) {};
		\node [style=none] (51) at (21, 3.75) {};
		\node [style=none] (52) at (21, -4) {};
		\node [style=none] (53) at (12.5, -4) {};
		\node [style=new style 0] (54) at (13.75, 1.75) {};
		\node [style=new style 0] (55) at (15.5, 2.5) {};
		\node [style=new style 0] (56) at (15, 1.25) {};
		\node [style=new style 0] (57) at (13.75, -2.75) {};
		\node [style=new style 0] (58) at (15.25, -1.75) {};
		\node [style=new style 0] (59) at (15.5, -3) {};
		\node [style=new style 0] (60) at (18, 1.5) {};
		\node [style=new style 0] (61) at (19, 2.5) {};
		\node [style=new style 0] (62) at (20, 2.5) {};
		\node [style=new style 0] (63) at (17.75, -0.5) {};
		\node [style=new style 0] (64) at (18.75, -1) {};
		\node [style=new style 0] (65) at (19.75, -1.25) {};
		\node [style=new style 0] (66) at (17.5, -2.5) {};
		\node [style=new style 0] (67) at (17.75, -3.25) {};
		\node [style=none] (86) at (3.5, 3) {};
		\node [style=none] (87) at (6.75, 3) {};
		\node [style=none] (88) at (6.75, -3.5) {};
		\node [style=none] (89) at (3.5, -3.5) {};
		\node [style=none] (90) at (7.5, -3.5) {};
		\node [style=none] (91) at (7.5, 3) {};
		\node [style=none] (92) at (11, 3) {};
		\node [style=none] (93) at (11, -3.5) {};
		\node [style=none] (94) at (-2.5, 3.75) {};
		\node [style=none] (95) at (-2.5, -4) {};
		\node [style=none] (96) at (3.5, -0.5) {};
		\node [style=none] (97) at (6.75, -0.5) {};
		\node [style=none] (98) at (7.5, 0.5) {};
		\node [style=none] (99) at (11, 0.5) {};
		\node [style=none] (100) at (13, 3) {};
		\node [style=none] (101) at (16, 3) {};
		\node [style=none] (102) at (16, 0.75) {};
		\node [style=none] (103) at (13, 0.75) {};
		\node [style=none] (104) at (13, -1) {};
		\node [style=none] (105) at (16, -1) {};
		\node [style=none] (106) at (16, -3.5) {};
		\node [style=none] (107) at (13, -3.5) {};
		\node [style=none] (108) at (17, 0) {};
		\node [style=none] (109) at (17, -3.75) {};
		\node [style=none] (110) at (20.25, -3.75) {};
		\node [style=none] (111) at (20.25, 0) {};
		\node [style=none] (112) at (17.5, 1) {};
		\node [style=none] (113) at (17.5, 3) {};
		\node [style=none] (114) at (20.25, 3) {};
		\node [style=none] (115) at (20.25, 1) {};
		\node [style=none] (116) at (17, -1) {};
		\node [style=none] (117) at (20.25, -2.75) {};
		\node [style=none] (118) at (22, 3.75) {};
		\node [style=none] (119) at (30.5, 3.75) {};
		\node [style=none] (120) at (30.5, -4) {};
		\node [style=none] (121) at (22, -4) {};
		\node [style=new style 0] (122) at (23.25, 1.75) {};
		\node [style=new style 0] (123) at (25, 2.5) {};
		\node [style=new style 0] (124) at (24.5, 1.25) {};
		\node [style=new style 0] (125) at (23.25, -2.75) {};
		\node [style=new style 0] (126) at (24.75, -1.75) {};
		\node [style=new style 0] (127) at (25, -3) {};
		\node [style=new style 0] (128) at (27.5, 1.5) {};
		\node [style=new style 0] (129) at (28.5, 2.5) {};
		\node [style=new style 0] (130) at (29.5, 2.5) {};
		\node [style=new style 0] (131) at (27.25, -0.5) {};
		\node [style=new style 0] (132) at (28.25, -1) {};
		\node [style=new style 0] (133) at (29.25, -1.25) {};
		\node [style=new style 0] (134) at (27, -2.5) {};
		\node [style=new style 0] (135) at (27.25, -3.25) {};
		\node [style=none] (136) at (22.5, 3) {};
		\node [style=none] (137) at (25.5, 3) {};
		\node [style=none] (138) at (25.5, 0.75) {};
		\node [style=none] (139) at (22.5, 0.75) {};
		\node [style=none] (140) at (22.5, -1) {};
		\node [style=none] (141) at (25.5, -1) {};
		\node [style=none] (142) at (25.5, -3.5) {};
		\node [style=none] (143) at (22.5, -3.5) {};
		\node [style=none] (144) at (26.5, 0) {};
		\node [style=none] (145) at (26.5, -3.75) {};
		\node [style=none] (146) at (29.75, -3.75) {};
		\node [style=none] (147) at (29.75, 0) {};
		\node [style=none] (148) at (27, 1) {};
		\node [style=none] (149) at (27, 3) {};
		\node [style=none] (150) at (29.75, 3) {};
		\node [style=none] (151) at (29.75, 1) {};
		\node [style=none] (152) at (26.5, -1.75) {};
		\node [style=none] (153) at (29.75, -1.75) {};
		\node [style=none] (154) at (26.5, -2) {};
		\node [style=none] (155) at (29.75, -2) {};
	\end{pgfonlayer}
	\begin{pgfonlayer}{edgelayer}
		\draw (14.center) to (15.center);
		\draw (15.center) to (16.center);
		\draw (17.center) to (16.center);
		\draw (17.center) to (14.center);
		\draw (32.center) to (33.center);
		\draw (33.center) to (34.center);
		\draw (35.center) to (34.center);
		\draw (35.center) to (32.center);
		\draw (50.center) to (51.center);
		\draw (51.center) to (52.center);
		\draw (53.center) to (52.center);
		\draw (53.center) to (50.center);
		\draw (86.center) to (87.center);
		\draw (87.center) to (88.center);
		\draw (88.center) to (89.center);
		\draw (89.center) to (86.center);
		\draw (91.center) to (92.center);
		\draw (92.center) to (93.center);
		\draw (91.center) to (90.center);
		\draw (90.center) to (93.center);
		\draw  [style=new edge style 0] (94.center) to (95.center);
		\draw  [style=new edge style 0] (96.center) to (97.center);
		\draw  [style=new edge style 0] (98.center) to (99.center);
		\draw (100.center) to (103.center);
		\draw (103.center) to (102.center);
		\draw (102.center) to (101.center);
		\draw (100.center) to (101.center);
		\draw (113.center) to (114.center);
		\draw (114.center) to (115.center);
		\draw (115.center) to (112.center);
		\draw (112.center) to (113.center);
		\draw (104.center) to (105.center);
		\draw (105.center) to (106.center);
		\draw (106.center) to (107.center);
		\draw (107.center) to (104.center);
		\draw (108.center) to (109.center);
		\draw (109.center) to (110.center);
		\draw (110.center) to (111.center);
		\draw (111.center) to (108.center);
		\draw [style=new edge style 0] (116.center) to (117.center);
		\draw (118.center) to (119.center);
		\draw (119.center) to (120.center);
		\draw (121.center) to (120.center);
		\draw (121.center) to (118.center);
		\draw (136.center) to (139.center);
		\draw (139.center) to (138.center);
		\draw (138.center) to (137.center);
		\draw (136.center) to (137.center);
		\draw (149.center) to (150.center);
		\draw (150.center) to (151.center);
		\draw (151.center) to (148.center);
		\draw (148.center) to (149.center);
		\draw (140.center) to (141.center);
		\draw (141.center) to (142.center);
		\draw (142.center) to (143.center);
		\draw (143.center) to (140.center);
		\draw (145.center) to (146.center);
		\draw (147.center) to (144.center);
		\draw (144.center) to (152.center);
		\draw (152.center) to (153.center);
		\draw (153.center) to (147.center);
		\draw (154.center) to (145.center);
		\draw (154.center) to (155.center);
		\draw (155.center) to (146.center);
	\end{pgfonlayer}
\end{tikzpicture}
\vspace{-0.2cm}
\caption{The hierarchical model selection procedure on a sample embedding. Dots represent vertex embeddings. Rectangles represent final or intermediate blocks. Dashed lines represent partitions returned by the $K$means algorithm.\label{fig:hierarchical}}
\vspace{-0.5cm}
\end{figure}
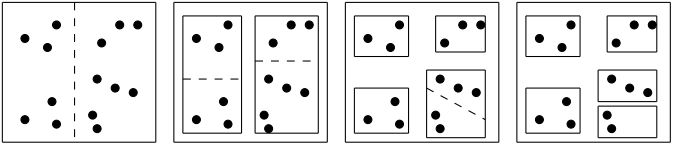

Our hierarchical model selection procedure based on the local modularity metric is summarized in \cref{alg:recursion}. We demonstrate it by an example (\cref{fig:hierarchical}). For a set of nodes $U \subseteq V$ to be partitioned, we first compute the local modularity $m_1$ that treats ${U}$ as a single block (i.e., Line \ref{line:singleblock}). The $K$Means clustering algorithm is then applied to embeddings $\{ {\bf{z}}_i | v_i \in U \}$ (i.e., Line \ref{line:partitionintotwo}), which tries to divide ${U}$ into two blocks $\{ U_1, U_2 \}$ based on the distances between embeddings. One can also compute the local modularity $m_2$ w.r.t. the partition of $\{ U_1, U_2 \}$ (i.e., Line \ref{line:twoblocks}). We then introduce a heuristic stopping rule based on the sizes of $\{ U_1 , U_2\}$ and values of $\{ m_1, m_2\}$. Concretely, we adopt the set $U$ as a new block, if $|U_1|$ or $|U_2|$ are less than or equal to a threshold $\epsilon$ (i.e., $\epsilon=5$ in our implementation) or $m_1 > m_2$ (i.e., Lines \ref{line:threshold},\ref{line:newblock}). Otherwise, we repeat the aforementioned procedures to further partition $U_1$ and $U_2$ (i.e., Lines \ref{line:recursion1}, \ref{line:recursion2}).

In particular, the stopping rule in Line \ref{line:threshold} can effectively avoid partitioning $U$ into too small or imbalanced blocks (e.g., with $|U_1| < \epsilon$ or $|U_2| < \epsilon$). Consistent with the modularity maximization objective, it also tries to avoid the decrease of modularity after the partitioning.

The complexity of this algorithm depends on the quality of the embedding. If the embedding is of good quality so that the $K$Means algorithm always partitions the node set $U$ into two balanced sets, we need to run the $K$Means algorithm on a total of $O(N\log N)$ nodes. The complexity in this case is $O(NT\log N)$ where $T$ is the number of iterations in $K$Means. In the worst case, the set $U$ is always partitioned into a single block and the remaining blocks. In this case, the $K$Means algorithm is run on $O(KN)$ nodes where $K$ is the number of blocks. The complexity in this case is $O(KTN)$.

\subsection{Extension to Streaming Graphs}

The aforementioned designs of RaftGP can be easily extended to streaming graphs.
Let $E'$ be the set of newly added edges in the streaming setting. Since the addition of $E'$ only changes the degrees of at most $2|E'|$ nodes, one can use an incremental strategy to update the input statistics ${\bf{X}} \in \{ {\bf{M}}, {\bf{\tilde Q}}\}$ with a complexity of $O(|E'|)$. Let $V'$ be the set of nodes incident to $E'$.
Note that only the multi-hop neighbors of $V'$ (denoted by $V''$) will participate in the GNN feature aggregation of $V'$. Let $E''$ be the set of edges induced by
$V' \cup V''$. One can also incrementally update embeddings $\{ {\bf{z}}_i \}$ via the random projection and GNN feedforward propagation induced by  $(V' \cup V'', E'')$. Therefore, the overall complexity of embedding derivation is $O(|E''|d + |V' \cup V''|d)$, where we usually have $|E''| < |E|$ and $|V' \cup V''| < N$.

For the hierarchical model selection, one can formulate the partitioning procedure as a tree, where each tree node represents a subset of nodes (i.e., $U \subseteq V$) to be partitioned; the children of this tree node represent a partition of $U$ (e.g., $(U_1, U_2)$); each leaf corresponds to a block $C_r$. The tree structure allows one to update only a tree path (path from a leaf to the root) once the embedding ${\bf{z}}_i$ of node $v_i$ is updated.

\section{Fast Graph Generating from the SBM}
\label{sec:data}\label{sec:sample}

The stochastic block partition benchmark of the IEEE HPEC Graph Challenge generates test graphs using the SBM generators (i.e., `\texttt{blockmodel}' and `\texttt{blockmodel-degree}') implemented by graph-tool \cite{peixoto_graph-tool_2014}.
We argue that there are some limitations in the implementations of these generators.

For instance, `\texttt{blockmodel}' ignores the degree correction terms (i.e., $\boldsymbol{\theta}$ in Definition~\ref{def:SBM}).
`\texttt{blockmodel-degree}' uses the Markov Chain Monte Carlo (MCMC) sampling algorithm to rewire the edges generated from another random model. As a result, the distribution $\mathcal{X}$ of its output approaches the SBM distribution $\mathcal{S}$ only after sufficient iterations of MCMC. In practice, the number of iterations is set to a constant. This makes the distribution $\mathcal{X}$ an approximation of $\mathcal{S}$ without a provable distance guarantee.
\begin{algorithm2e}[t]\small
\caption{\small Fast Graph Generating from the SBM}
\label[Algorithm]{alg:sample}
\SetKwProg{Proc}{procedure}{}{}
\Proc{$\textsc{RangeSum}(r, s, x, y)$} {
    \Return Sum of $\{ \lambda_{ij} \equiv \boldsymbol{\theta}_i \boldsymbol{\theta}_j {\bf{\Omega}}_{c_i c_j} \}$ over the $x$-th to the $y$-th edge between blocks $C_r$ and $C_s$
}
\Proc{$\textsc{Sample}(N, K, {\bf{c}}, \boldsymbol{\theta}, \OOmega)$}{
        Let $C_r\leftarrow \{v_i | {\bf{c}}_i = r\}$ 
        be the $r$-th block\\
    \For{all pairs $(r, s)$ {\bf{from}} $[K]\times [K]$}
    {
            $m\leftarrow 1$ // Counter of edge\\
            \While{$m \le |C_r| |C_s|$} {
                Sample $p \sim U[0,1]$\\
                Let $t$ be the smallest integer in $[m,|C_r||C_s|]$ s.t. $p > \exp {(- \textsc{RangeSum}(r, s, m, t))}$\label{line:computex}\\
                \If {$t$ does not exist\label{line:break}} {
                    \Break\label{line:b}\\
                }
                Let $(v_i, v_j)$ be the $t$-th edge from $C_r$ to $C_s$\\
                Generate $\AA_{ij} \sim Pois(\lambda_{ij})$ s.t. $\AA_{ij} \ne 0$\label{line:poisson}\\
                $m \leftarrow t+1$
            }
    }
}
\end{algorithm2e}

To address the limitations, we introduce a new fast graph generating algorithm. It generates graphs from the exact SBM distribution $\mathcal{S}(\boldsymbol{\theta}, \bf{c}, \OOmega)$ and has a nearly linear time complexity of $O((|E| + K^2) \log N)$, where $\{ \boldsymbol{\theta}, \bf{c}, \OOmega \}$ are defined in Definition~\ref{def:SBM}; $K$, $N$, and $|E|$ are the numbers of blocks, nodes, and edges in the generated graph.

\cref{alg:sample} summarizes the graph generating procedure, where we enumerate all pairs of blocks $(C_r, C_s)$ ($r \le s$) and sample edges between these two blocks. The algorithm repeatedly finds the next edge with nonzero weight (Lines \ref{line:computex} and \ref{line:break}) and determines its weight from a Poisson distribution (Line \ref{line:poisson}). To implement Line \ref{line:computex}, we perform a binary search on $t$. To compute $\textsc{RangeSum}(r, s, x, y)$, we arrange all the possible edges between $(C_r, C_s)$ in a virtual $|C_r|\times |C_s|$ table. Then we can see that a precomputed 2D prefix sum gives $\textsc{RangeSum}(r, s, x, y)$ in $O(1)$ time. Note that for undirected graphs, we should ignore the lower diagonal entries in $\AA$.

\section{Experiments}\label{sec:experiment}

\subsection{Experiment Setups}

\subsubsection{\textbf{Datasets}}

\begin{table}[t]\scriptsize 
\caption{Details of the Generated Benchmark Datasets}\label{tab:data}
\vspace{-0.2cm}
\centering
\begin{tabular}{l|l|l|l|l|l}
\hline
$N$ & $|E|$ & $K$ & deg & density & \textbf{Layer Configuration}\\ \hline
1E3 &  1.69E4$\sim$1.74E4 & 11 & 9$\sim$112 & 3E-2 & [256,128,64]\\
5E3 &  9.51E4$\sim$9.69E4 & 19 & 5$\sim$164 & 8E-3 & [1024,512,256,128]\\
1E4 &  1.95E5$\sim$1.96E5 & 25 & 6$\sim$180 & 4E-3 & [4096,2048,1024,512,256]\\
5E4 &  9.97E5$\sim$1.01E6 & 44 & 5$\sim$205 & 8E-4 & [4096,2048,1024,512,256]\\
1E5 &  2.01E6$\sim$2.02E6 & 56 & 4$\sim$209 & 4E-4 & [4096,2048,1024,512,256]\\
2E5 &  4.04E6$\sim$4.06E6 & 71 & 5$\sim$230 & 2E-4 & [4096,2048,1024,512,256]\\ \hline
\end{tabular}
\vspace{-0.5cm}
\end{table}

We followed the stochastic block partitioning benchmark of the IEEE HPEC Graph Challenge to generate test datasets from the SBM. As mentioned in Section~\ref{sec:data}, we develop a new generator to overcome the limitations of the original implementations in the benchmark.

We set (\romannumeral1) the ratio between the numbers of within-block edges and between-block edges to $2.5$ and (\romannumeral2) block size heterogeneity to $3$. These configurations correspond to the hardest setting of the benchmark.
We generated graphs with different scales by respectively setting the number of nodes $N$ to 1E3, 5E3, 1E4, 5E4, 1E5, and 2E5. For each setting of $N$, we independently generated five graphs.
Statistics of the generated datasets are shown in Table~\ref{tab:data}, where $|E|$ and $K$ are the numbers of edges and blocks; ${\deg}$ is the node degree.

\subsubsection{\textbf{Baselines}}
We compared RaftGP with five baselines, including \textit{MC-SBM} \cite{tiago14}, \textit{greedy modularity maximization} (\textit{GMod}) \cite{clauset2004finding}, \textit{spectral modularity maximization} (\textit{SMod}) \cite{newman2006modularity}, \textit{Par-SBM} \cite{peng2017scalable}, and \textit{BP-Mod} \cite{zhang2014scalable}.
As stated in Definition~\ref{def:GP}, we consider the GP task where the number of blocks $K$ is unknown and determined by the method to be evaluated. Hence, some other baselines that need a pre-set $K$ (e.g., \textit{metis} \cite{hendrickson1995multi}, \textit{GraClus} \cite{dhillon2007weighted}, and \textit{spectral clustering} \cite{von2007tutorial}) could not be included in our experiments.

Note that RaftGP has two variants with their features extracted from NCut minimization (i.e., ${\bf{X}} = {\bf{M}}$) and modularity maximization (i.e., ${\bf{X}} = {\bf{\tilde Q}}$).They are denoted as \textit{RaftGP-C} and \textit{RaftGP-M}, respectively.

\subsubsection{\textbf{Evaluation Criteria}}
We followed the IEEE HPEC Graph Challenge to adopt \textit{accuracy}, \textit{adjusted random index} (ARI), \textit{precision}, and \textit{recall} as quality metrics. Given precision and recall, we also computed the corresponding \textit{F1-score} as a comprehensive metric considering both aspects.
The \textit{runtime} (in terms of seconds) of a method to output its GP result was used as the efficiency metric.
We also define that a method encounters the out-of-time (OOT) exception if it cannot obtain a GP result within 1E4 seconds.

\subsubsection{\textbf{Parameter \& Environment Settings}}

Layer configurations of RaftGP on all the datasets are shown in Table~\ref{tab:data}, where the first and last numbers denote the input feature dimensionality $L$ and embedding dimensionality $d$.
We used Python 3.7 and PyTorch 1.13 to implement RaftGP and adopted the official or widely-used implementations of baselines (implemented by C++ and Python). All the experiments were conducted on a server with two Intel Xeon 6130 CPUs (16 cores), one GeForce RTX3090 GPU (24GB GPU memory), 32GB main memory, and Ubuntu 18.04.5 LTS OS.

\subsection{Quantitative Evaluation \& Discussions}

\begin{table*}[t] 
\caption{Quantitative Evaluation Results}\label{tab:eva}
\vspace{-0.2cm}
\centering
\begin{tabular}{c|l|p{0.5cm}p{0.5cm}l|p{0.95cm}|c|l|p{0.5cm}p{0.5cm}l|p{0.9cm}}
\hline
\multicolumn{1}{l|}{$N$} & \textbf{Methods} & \textbf{Acc}$\uparrow$ & \textbf{ARI}$\uparrow$ & \textbf{F1}$\uparrow$ (Recall, Precision) & \textbf{Time}$\downarrow$(s) & \multicolumn{1}{l|}{$N$} & \textbf{Methods} & \textbf{Acc}$\uparrow$ & \textbf{ARI}$\uparrow$ & \textbf{F1}$\uparrow$ (Recall, Precision) & \textbf{Time}$\downarrow$(s) \\ \hline
\multirow{7}{*}{1E3} & MC-SBM & 0.9764 & 0.9689 & 0.9736 (0.9486, 1.0000) & 10.4589 & \multirow{7}{*}{5E4} & MC-SBM & 0.9769 & 0.9701 & 0.9711 (0.9439, 1.0000) & 1658.27 \\
 & GMod & 0.7098 & 0.6813 & 0.7309 (0.9823, 0.5820) & 4.7274 &  & GMod & OOT & OOT & OOT & $>$10,000 \\
 & SMod & 0.7098 & 0.5879 & 0.6531 (0.8675, 0.5237) & 42.4847 &  & SMod & OOM & OOM & OOM & OOM \\
 & Par-SBM & 0.9984 & 0.9976 & 0.9978 (0.9959, 0.9998) & 0.4022 &  & Par-SBM & 0.9748 & 0.9747 & 0.9758 (0.9998, 0.9529) & 14.2388 \\
 & BP-SBM & 0.6904 & 0.6924 & 0.7508 (0.9999, 0.6010) & 261.7399 &  & BP-SBM & OOT & OOT & OOT & $>$10,000 \\
 & \textbf{\textit{RaftGP-C}} & \textbf{0.9996} & \textbf{0.9994} & \textbf{0.9994} (0.9997, 0.9992) & 1.6172 &  & \textbf{\textit{RaftGP-C}} & \textbf{0.9998} & \textbf{0.9998} & \textbf{0.9998} (1.0000, 0.9996) & 71.1237 \\
 & \textbf{\textit{RaftGP-M}} & \textbf{0.9996} & \textbf{0.9994} & \textbf{0.9994} (0.9997, 0.9992) & 1.5634 &  & \textbf{\textit{RaftGP-M}} & \textbf{0.9998} & \textbf{0.9998} & \textbf{0.9998} (1.0000, 0.9996) & 69.4147 \\ \hline
\multirow{7}{*}{5E3} & MC-SBM & 0.9841 & 0.9858 & 0.9868 (0.9739, 1.0000) & 83.0489 & \multirow{7}{*}{1E5} & MC-SBM & OOT & OOT & OOT & $>$10,000 \\
 & GMod & 0.6172 & 0.5930 & 0.6289 (0.9627, 0.4670) & 120.4837 &  & GMod & OOT & OOT & OOT & $>$10,000 \\
 & SMod & 0.3639 & 0.1939 & 0.2848 (0.7149, 0.1778) & 1234.89 &  & SMod & OOM & OOM & OOM & OOM \\
 & Par-SBM & 0.9916 & 0.9894 & 0.9904 (0.9984, 0.9825) & 1.1748 &  & Par-SBM & 0.9061 & 0.9096 & 0.9130 (0.9998, 0.8401) & 38.9002 \\
 & BP-SBM & 0.4746 & 0.4755 & 0.5282 (0.9999, 0.3589) & 3994.94 &  & BP-SBM & OOT & OOT & OOT & $>$10,000 \\
 & \textbf{\textbf{\textit{RaftGP-C}}} & \textbf{0.9998} & \textbf{0.9998} & \textbf{0.9997} (0.9999, 0.9996) & 5.9843 &  & \textbf{\textbf{\textit{RaftGP-C}}} & \textbf{1.0000} & \textbf{1.0000} & \textbf{1.0000} (1.0000, 1.0000) & 150.6443 \\
 & \textbf{\textit{RaftGP-M}} & \textbf{0.9998} & \textbf{0.9998} & \textbf{0.9997} (0.9999, 0.9996) & 6.1395 &  & \textbf{\textit{RaftGP-M}} & \textbf{1.0000} & \textbf{1.0000} & \textbf{1.0000} (1.0000, 1.0000) & 154.8510 \\ \hline
\multirow{7}{*}{1E4} & MC-SBM & 0.9608 & 0.9495 & 0.9539 (0.9119, 0.9999) & 313.1485 & \multirow{7}{*}{2E5} & MC-SBM & OOT & OOT & OOT & $>$10,000 \\
 & GMod & OOT & OOT & OOT & $>$10,000 &  & GMod & OOT & OOT & OOT & $>$10,000 \\
 & SMod & 0.4693 & 0.3066 & 0.3721 (0.7592, 0.2464) & 6609.61 &  & SMod & OOM & OOM & OOM & OOM \\
 & Par-SBM & 0.9908 & 0.9959 & 0.9962 (0.9992, 0.9933) & 4.4148 &  & Par-SBM & 0.9341 & 0.9445 & 0.9461 (0.9999, 0.8978) & 109.4829 \\
 & BP-SBM & OOT & OOT & OOT & $>$10,000 &  & BP-SBM & OOT & OOT & OOT & $>$10,000 \\
 & \textbf{\textbf{\textit{RaftGP-C}}} & \textbf{0.9981} & \textbf{0.9974} & \textbf{0.9975} (0.9999, 0.9952) & 15.6803 &  & \textbf{\textbf{\textit{RaftGP-C}}} & \textbf{1.0000} & \textbf{1.0000} & \textbf{1.0000} (1.0000, 1.0000) & 377.0465 \\
 & \textbf{\textit{RaftGP-M}} & \textbf{0.9981} & 0.9966 & 0.9968 (0.9999, 0.9938) & 15.6424 &  & \textbf{\textit{RaftGP-M}} & \textbf{1.0000} & \textbf{1.0000} & \textbf{1.0000} (1.0000, 1.0000) & 367.5187 \\ \hline
\end{tabular}
\vspace{-0.4cm}
\end{table*}

\begin{table}[t] 
\caption{Detailed Runtime of RaftGP in terms of Seconds}\label{tab:time}
\vspace{-0.1cm}
\centering
\begin{tabular}{c|l|l|lll}
\hline
\multicolumn{1}{l|}{$N$} & \textbf{Methods} & \textbf{Total} (s) & \textbf{Feat} & \textbf{Emb} & \textbf{Model} \\ \hline
\multirow{2}{*}{1E3} & \textbf{\textit{RaftGP-C}} & 1.6172 & 0.1172 & 0.2454 & 1.2547 \\
 & \textbf{\textit{RaftGP-M}} & 1.5634 & 0.1136 & 0.2320 & 1.2177 \\ \hline
\multirow{2}{*}{5E3} & \textbf{\textit{RaftGP-C}} & 5.9843 & 0.5228 & 0.2525 & 5.2089 \\
 & \textbf{\textit{RaftGP-M}} & 6.1395 & 0.4374 & 0.2537 & 5.4484 \\ \hline
\multirow{2}{*}{1E4} & \textbf{\textit{RaftGP-C}} & 15.6803 & 2.7233 & 0.3040 & 12.6531 \\
 & \textbf{\textit{RaftGP-M}} & 15.6424 & 2.4969 & 0.2788 & 12.8667 \\ \hline
\multirow{2}{*}{5E4} & \textbf{\textit{RaftGP-C}} & 71.1237 & 11.2939 & 0.6683 & 59.1615 \\
 & \textbf{\textit{RaftGP-M}} & 69.4147 & 10.4094 & 0.4556 & 58.5497 \\ \hline
\multirow{2}{*}{1E5} & \textbf{\textit{RaftGP-C}} & 150.6443 & 23.1460 & 0.7190 & 126.7793 \\
 & \textbf{\textit{RaftGP-M}} & 154.8510 & 24.0318 & 0.7432 & 130.0760 \\ \hline
\multirow{2}{*}{2E5} & \textbf{\textit{RaftGP-C}} & 377.0465 & 62.1050 & 1.3236 & 313.6178 \\
 & \textbf{\textit{RaftGP-M}} & 367.5187 & 57.6826 & 1.5124 & 308.3236 \\ \hline
\end{tabular}
\vspace{-0.4cm}
\end{table}

The average quality metrics and runtime over five generated graphs w.r.t. each setting of $N$ are reported in Table~\ref{tab:eva}, where the quality metrics of RaftGP are in \textbf{bold} if they perform the best; OOT and OOM denote the out-of-time and out-of-memory exceptions.
In addition to the total runtime, Table~\ref{tab:time} reports the time of (\romannumeral1) random feature extraction, (\romannumeral2) embedding derivation (i.e., GNN feedforward propagation), and (\romannumeral3) model selection for RaftGP.

Note that RaftGP does not include any training procedures. Surprisingly, both variants of RaftGP can achieve the best quality metrics in most cases, significantly outperforming \textit{MC-SBM}, \textit{GMod}, \textit{SMod}, and \textit{BP-SBM}. It verifies our discussions in Section~\ref{sec:emb-der} that \textit{one feedforward propagation through a random initialized GNN is a special random projection with the geometric properties of the aggregated features preserved, and thus can still derive informative embeddings}.
Although \textit{Par-SBM} has close quality and slightly faster runtime compared with RaftGP when $N$ is small, our methods can achieve much better quality metrics in cases with larger $N$s. In this sense, we believe that \textit{RaftGP can achieve a better trade-off between quality and efficiency especially when $N$ is large}.

On all the datasets, the two variants of RaftGP have close quality metrics with best performance. It validates our motivation of \textit{using ${\bf{M}}$ and ${\bf{\tilde Q}}$ (extracted from the NCut minimization and modularity maximization objectives) as informative statistics for the derivation of community-preserving embeddings}.

As in Table~\ref{tab:time}, model selection is the major bottleneck for both variants of RaftGP, which accounts for more than $80$\% of the total runtime.
It implies that \textit{the random feature extraction and GNN-based embedding derivation are efficient designs for RaftGP} with runtime better than \textit{Par-SBM}.

\subsection{Ablation Study}

\begin{table}[t] 
\caption{Ablation Study of RaftGP}\label{tab:abl}
\vspace{-0.1cm}
\centering
\begin{tabular}{c|l|lll}
\hline
\multicolumn{1}{l|}{$N$} & \textbf{Methods} & \textbf{Acc}$\uparrow$ & \textbf{ARI}$\uparrow$ & \textbf{F1}$\uparrow$ (Recall, Precision) \\ \hline
\multirow{6}{*}{1E3} & \textit{\textbf{RaftGP-C}} & \textbf{0.9996} & \textbf{0.9994} & \textbf{0.9994} (0.9997, 0.9992) \\
 & ~(\romannumeral1) w/o Feat & 0.9996 & 0.9994 & 0.9994 (0.9997, 0.9992) \\
 & ~(\romannumeral2)w/o Emb & 0.9806 & 0.9903 & 0.9916 (0.9997, 0.9836) \\ \cline{2-5} 
 & \textit{\textbf{RaftGP-M}} & \textbf{0.9996} & \textbf{0.9994} & \textbf{0.9994} (0.9997, 0.9992) \\
 & ~(\romannumeral1) w/o Feat & 0.9996 & 0.9994 & 0.9994 (0.9997, 0.9992) \\
 & ~(\romannumeral2)w/o Emb & 0.9000 & 0.8936 & 0.9190 (0.9998, 0.8502) \\ \hline
\multirow{6}{*}{5E3} & \textit{\textbf{RaftGP-C}} & \textbf{0.9998} & \textbf{0.9998} & \textbf{0.9997} (0.9999, 0.9996) \\
 & ~(\romannumeral1) w/o Feat & 0.9998 & 0.9998 & 0.9997 (0.9999, 0.9996) \\
 & ~(\romannumeral2)w/o Emb & 0.9846 & 0.9784 & 0.9798 (0.9999, 0.9605) \\ \cline{2-5} 
 & \textit{\textbf{RaftGP-M}} & \textbf{0.9998} & \textbf{0.9998} & \textbf{0.9997} (0.9999, 0.9996) \\
 & ~(\romannumeral1) w/o Feat & 0.9998 & 0.9998 & 0.9997 (0.9999, 0.9996) \\
 & ~(\romannumeral2)w/o Emb & 0.1257 & 0.0067 & 0.1266 (0.9996, 0.0676) \\ \hline
\multirow{6}{*}{1E4} & \textit{\textbf{RaftGP-C}} & \textbf{0.9981} & \textbf{0.9974} & \textbf{0.9975} (0.9999, 0.9952) \\
 & ~(\romannumeral1) w/o Feat & 0.9981 & 0.9967 & 0.9968 (0.9999, 0.9938) \\
 & ~(\romannumeral2)w/o Emb & 0.8452 & 0.7023 & 0.7534 (0.9998, 0.6045) \\ \cline{2-5} 
 & \textit{\textbf{RaftGP-M}} & \textbf{0.9981} & \textbf{0.9966} & \textbf{0.9968} (0.9999, 0.9938) \\
 & ~(\romannumeral1) w/o Feat & 0.9981 & 0.9966 & 0.9968 (0.9999, 0.9938) \\
 & ~(\romannumeral2)w/o Emb & 0.2725 & 0.0914 & 0.1966 (1.0000, 0.1090) \\ \hline
\multirow{6}{*}{5E4} & \textit{\textbf{RaftGP-C}} & \textbf{0.9998} & \textbf{0.9998} & \textbf{0.9998} (1.0000, 0.9996) \\
 & ~(\romannumeral1) w/o Feat & OOM & OOM & OOM \\
 & ~(\romannumeral2)w/o Emb & 0.0537 & 0.0000 & 0.0543 (1.0000, 0.0279) \\ \cline{2-5} 
 & \textit{\textbf{RaftGP-M}} & \textbf{0.9998} & \textbf{0.9998} & \textbf{0.9998} (1.0000, 0.9996) \\
 & ~(\romannumeral1) w/o Feat & OOM & OOM & OOM \\
 & ~(\romannumeral2)w/o Emb & 0.0829 & 0.0261 & 0.0791 (0.9998, 0.0412) \\ \hline
\end{tabular}
\vspace{-0.5cm}
\end{table}

We also validated the effects of (\romannumeral1) random feature extraction and (\romannumeral2) embedding derivation (with GNN) of RaftGP by removing the corresponding components. In case (\romannumeral1), ${\bf{X}} \in \{ {\bf{M}}, {\bf{\tilde Q}} \}$ was directly used as the input of GNN (without random projection). In case (\romannumeral2), we directly treated the features ${\bf{Y}}$ derived via random projection as the embedding input of model selection (without using GNN).
Results of our ablation study on all the datasets are shown in Table~\ref{tab:abl}.

Although case (\romannumeral1) can achieve the quality competitive to the original RaftGP model when $N$ is small, we encounter the OOM exception when $N \ge$ 5E4. It indicates that \textit{the Gaussian random projection can reduce the feature dimensionality while preserving the key geometric properties of input features} as discussed in Section~\ref{sec:feat_ext}.
Moreover, the GNN-based embedding derivation is essential to ensure the high quality of RaftGP because there are significant quality declines in case (\romannumeral2) with the increase of $N$. It verifies our discussions in Section~\ref{sec:emb-der} that \textit{the feature aggregation of GNN can enhance the ability of RaftGP to preserve community structures}.

\section{Conclusion}\label{sec:conclusion}
In this paper, we focused on the GP problem and proposed a RaftGP method based on an efficient community-preserving embedding scheme, including the random feature extraction, random embedding derivation, and hierarchical model selection.
We found that a randomly initialized GNN, even without training, is enough to derive informative community-preserving embeddings and support high-quality GP.
We evaluated our method on the stochastic block partitioning benchmark of the IEEE HPEC Graph Challenge and compared the performance with five baselines, where a new graph generating algorithm was developed to address some limitations of the original data generators in the benchmark. Compared to the baseline provided by the IEEE HPEC Graph Challenge, our method achieves better accuracy on all test cases and takes 64.5x shorter time on large graphs.
In our future work, we plan to extend RaftGP to dynamic graphs \cite{lei2018adaptive,lei2019gcn,qin2022temporal,qin2023high}.


\bibliographystyle{IEEEtran}

\begin{thebibliography}{10}
\providecommand{\url}[1]{#1}
\csname url@samestyle\endcsname
\providecommand{\newblock}{\relax}
\providecommand{\bibinfo}[2]{#2}
\providecommand{\BIBentrySTDinterwordspacing}{\spaceskip=0pt\relax}
\providecommand{\BIBentryALTinterwordstretchfactor}{4}
\providecommand{\BIBentryALTinterwordspacing}{\spaceskip=\fontdimen2\font plus
\BIBentryALTinterwordstretchfactor\fontdimen3\font minus \fontdimen4\font\relax}
\providecommand{\BIBforeignlanguage}[2]{{%
\expandafter\ifx\csname l@#1\endcsname\relax
\typeout{** WARNING: IEEEtran.bst: No hyphenation pattern has been}%
\typeout{** loaded for the language `#1'. Using the pattern for}%
\typeout{** the default language instead.}%
\else
\language=\csname l@#1\endcsname
\fi
#2}}
\providecommand{\BIBdecl}{\relax}
\BIBdecl

\bibitem{tiago14}
T.~Peixoto, ``Efficient monte carlo and greedy heuristic for the inference of stochastic block models,'' \emph{Physical Review. E, Statistical, Nonlinear, \& Soft Matter Physics}, vol.~89, p. 012804, 01 2014.

\bibitem{qin2018adaptive}
M.~Qin, D.~Jin, K.~Lei, B.~Gabrys, and K.~Musial-Gabrys, ``Adaptive community detection incorporating topology and content in social networks,'' \emph{Knowledge-Based Systems}, vol. 161, pp. 342--356, 2018.

\bibitem{li2019identifying}
W.~Li, M.~Qin, and K.~Lei, ``Identifying interpretable link communities with user interactions and messages in social networks,'' in \emph{Proceedings of the 2019 IEEE International Conference on Parallel \& Distributed Processing with Applications, Big Data \& Cloud Computing, Sustainable Computing \& Communications, Social Computing \& Networking (ISPA/BDCloud/SocialCom/SustainCom)}.\hskip 1em plus 0.5em minus 0.4em\relax IEEE, 2019, pp. 271--278.

\bibitem{qin2021dual}
M.~Qin and K.~Lei, ``Dual-channel hybrid community detection in attributed networks,'' \emph{Information Sciences}, vol. 551, pp. 146--167, 2021.

\bibitem{hendrickson2000graph}
B.~Hendrickson and T.~G. Kolda, ``Graph partitioning models for parallel computing,'' \emph{Parallel Computing}, vol.~26, no.~12, pp. 1519--1534, 2000.

\bibitem{dai2017optimal}
L.~Dai and B.~Bai, ``Optimal decomposition for large-scale infrastructure-based wireless networks,'' \emph{IEEE Transactions on Wireless Communications}, vol.~16, no.~8, pp. 4956--4969, 2017.

\bibitem{qin2019towards}
M.~Qin, K.~Lei, B.~Bai, and G.~Zhang, ``Towards a profiling view for unsupervised traffic classification by exploring the statistic features and link patterns,'' in \emph{Proceedings of the 2019 ACM SIGCOMM NetAI Workshop}, 2019, pp. 50--56.

\bibitem{DBLP:conf/hpec/KaoGHJKMMRSSSS17}
E.~K. Kao, V.~Gadepally, M.~B. Hurley, M.~Jones, J.~Kepner, S.~Mohindra, P.~Monticciolo, A.~Reuther, S.~Samsi, W.~Song, D.~Staheli, and S.~T. Smith, ``Streaming graph challenge: Stochastic block partition,'' in \emph{Proceedings of the 2017 IEEE High Performance Extreme Computing Conference (HPEC)}, 2017, pp. 1--12.

\bibitem{zhuzhunashvili2017preconditioned}
D.~Zhuzhunashvili and A.~Knyazev, ``Preconditioned spectral clustering for stochastic block partition streaming graph challenge (preliminary version at arxiv.),'' in \emph{Proceedings of the 2017 IEEE High Performance Extreme Computing Conference (HPEC)}.\hskip 1em plus 0.5em minus 0.4em\relax IEEE, 2017, pp. 1--6.

\bibitem{durbeck2021dpgs}
L.~Durbeck and P.~Athanas, ``Dpgs graph summarization preserves community structure,'' in \emph{Proceedings of the 2021 IEEE High Performance Extreme Computing Conference (HPEC)}.\hskip 1em plus 0.5em minus 0.4em\relax IEEE, 2021, pp. 1--9.

\bibitem{uppal2021faster}
A.~J. Uppal, J.~Choi, T.~B. Rolinger, and H.~H. Huang, ``Faster stochastic block partition using aggressive initial merging, compressed representation, and parallelism control,'' in \emph{Proceedings of the 2021 IEEE High Performance Extreme Computing Conference (HPEC)}.\hskip 1em plus 0.5em minus 0.4em\relax IEEE, 2021, pp. 1--7.

\bibitem{wanye2019fast}
F.~Wanye, V.~Gleyzer, and W.-c. Feng, ``Fast stochastic block partitioning via sampling,'' in \emph{Proceedings of the 2019 IEEE High Performance Extreme Computing Conference (HPEC)}.\hskip 1em plus 0.5em minus 0.4em\relax IEEE, 2019, pp. 1--7.

\bibitem{zhang2020deep}
Z.~Zhang, P.~Cui, and W.~Zhu, ``Deep learning on graphs: A survey,'' \emph{IEEE Transactions on Knowledge \& Data Engineering (TKDE)}, vol.~34, no.~1, pp. 249--270, 2020.

\bibitem{arriaga2006algorithmic}
R.~I. Arriaga and S.~Vempala, ``An algorithmic theory of learning: Robust concepts and random projection,'' \emph{Machine Learning}, vol.~63, pp. 161--182, 2006.

\bibitem{von2007tutorial}
U.~Von~Luxburg, ``A tutorial on spectral clustering,'' \emph{Statistics \& Computing}, vol.~17, no.~4, pp. 395--416, 2007.

\bibitem{newman2006modularity}
M.~E. Newman, ``Modularity and community structure in networks,'' \emph{Proceedings of the National Academy of Sciences (PNAS)}, vol. 103, no.~23, pp. 8577--8582, 2006.

\bibitem{wilder2019end}
B.~Wilder, E.~Ewing, B.~Dilkina, and M.~Tambe, ``End to end learning and optimization on graphs,'' \emph{Proceedings of the 2019 Advances in Neural Information Processing Systems (NIPS)}, vol.~32, 2019.

\bibitem{chen2019supervised}
Z.~Chen, J.~Bruna, and L.~Li, ``Supervised community detection with line graph neural networks,'' in \emph{Proceedings of the 7th International Conference on Learning Representations (ICLR)}, 2019.

\bibitem{qin2022trading}
M.~Qin, C.~Zhang, B.~Bai, G.~Zhang, and D.-Y. Yeung, ``Towards a better trade-off between quality and efficiency of community detection: An inductive embedding method across graphs,'' \emph{ACM Transactions on Knowledge Discovery from Data (TKDD)}, 2023.

\bibitem{tian2014learning}
F.~Tian, B.~Gao, Q.~Cui, E.~Chen, and T.-Y. Liu, ``Learning deep representations for graph clustering,'' in \emph{Proceedings of the 28th AAAI Conference on Artificial Intelligence}, vol.~28, no.~1, 2014.

\bibitem{yang2016modularity}
L.~Yang, X.~Cao, D.~He, C.~Wang, X.~Wang, and W.~Zhang, ``Modularity based community detection with deep learning,'' in \emph{Proceedings of the 25th International Joint Conference on Artificial Intelligence (IJCAI)}, vol.~16, 2016, pp. 2252--2258.

\bibitem{Bourgain1985embedding}
J.~Bourgain, ``On lipschitz embedding of finite metric spaces in hilbert space,'' \emph{Israel Journal of Mathematics}, vol.~52, no. 1-2, pp. 46--52, 1985.

\bibitem{1984Extensions}
W.~J.~J. Lindenstrauss, ``Extensions of lipschitz maps into a hilbert space,'' \emph{Contemporary Mathematics}, vol.~26, no. 189, pp. 189--206, 1984.

\bibitem{kipf2016semi}
T.~N. Kipf and M.~Welling, ``Semi-supervised classification with graph convolutional networks,'' in \emph{Proceedings of the 5th International Conference on Learning Representations (ICLR)}, 2017, p. 1609.02907.

\bibitem{peixoto_graph-tool_2014}
\BIBentryALTinterwordspacing
T.~P. Peixoto, ``The graph-tool python library,'' \emph{figshare}, 2014. [Online]. Available: \url{http://figshare.com/articles/graph_tool/1164194}
\BIBentrySTDinterwordspacing

\bibitem{clauset2004finding}
A.~Clauset, M.~E. Newman, and C.~Moore, ``Finding community structure in very large networks,'' \emph{Physical Review E}, vol.~70, no.~6, p. 066111, 2004.

\bibitem{peng2017scalable}
C.~Peng, Z.~Zhang, K.-C. Wong, X.~Zhang, and D.~E. Keyes, ``A scalable community detection algorithm for large graphs using stochastic block models,'' \emph{Intelligent Data Analysis}, vol.~21, no.~6, pp. 1463--1485, 2017.

\bibitem{zhang2014scalable}
P.~Zhang and C.~Moore, ``Scalable detection of statistically significant communities and hierarchies, using message passing for modularity,'' \emph{Proceedings of the National Academy of Sciences (PNAS)}, vol. 111, no.~51, pp. 18\,144--18\,149, 2014.

\bibitem{hendrickson1995multi}
B.~Hendrickson, R.~W. Leland \emph{et~al.}, ``A multi-level algorithm for partitioning graphs,'' \emph{Proceedings Supercomputing}, vol.~95, no.~28, pp. 1--14, 1995.

\bibitem{dhillon2007weighted}
I.~S. Dhillon, Y.~Guan, and B.~Kulis, ``Weighted graph cuts without eigenvectors a multilevel approach,'' \emph{IEEE Transactions on Pattern Analysis \& Machine Intelligence (TPAMI)}, vol.~29, no.~11, pp. 1944--1957, 2007.

\bibitem{lei2018adaptive}
K.~Lei, M.~Qin, B.~Bai, and G.~Zhang, ``Adaptive multiple non-negative matrix factorization for temporal link prediction in dynamic networks,'' in \emph{Proceedings of the 2018 SIGCOMM NetAI Workshop}, 2018, pp. 28--34.

\bibitem{lei2019gcn}
K.~Lei, M.~Qin, B.~Bai, G.~Zhang, and M.~Yang, ``Gcn-gan: A non-linear temporal link prediction model for weighted dynamic networks,'' in \emph{Proceedings of the 2019 IEEE Conference on Computer Communications (INFOCOM)}.\hskip 1em plus 0.5em minus 0.4em\relax IEEE, 2019, pp. 388--396.

\bibitem{qin2022temporal}
M.~Qin and D.-Y. Yeung, ``Temporal link prediction: A unified framework, taxonomy, and review,'' \emph{arXiv preprint arXiv:2210.08765}, 2022.

\bibitem{qin2023high}
M.~Qin, C.~Zhang, B.~Bai, G.~Zhang, and D.-Y. Yeung, ``High-quality temporal link prediction for weighted dynamic graphs via inductive embedding aggregation,'' \emph{IEEE Transactions on Knowledge \& Data Engineering (TKDE)}, vol.~35, no.~9, pp. 9378--9393, 2023.

\end{thebibliography}

\end{document}